# Exactly Solvable Sextic Potential Having Symmetric Triple-Well Structure

*Benbourenane Jamal[1*], Benbourenane Mohamed[1], and Eleuch Hichem[2,3]*

*Corresponding Author: E-mail:jamalben@hotmail.com

**Abstract**

In this paper, we introduce a family of sextic potentials that are exactly solvable, and for the first time, a family of triple-well potentials with their whole energy spectrum and wavefunctions using supersymmetry method.

It was suggested since three decades ago that all "additive" or "translational" shape invariant superpotentials formed by two combination of functions have been found and their list was already exhausted by the well-known exactly solvable potentials that are available in most textbooks and furthermore, there are no others.

We have devised a new family of superpotentials formed by a linear combination of three functions (two monomials and one rational) and where the change of parameter function is linear in four parameters. This new family of potentials with superpotential $W(x, A, B, D, G) = Ax^3 + Bx - \frac{Dx}{1+Gx^2}$ will extend the list of exactly solvable Schrödinger equations.

We have shown that the energy of the bound states is rational in the quantum number. Furthermore, approximating the potential around the central well by a harmonic oscillator, as a usual practice, is not valid. The two outer wells affect noticeably the probability density distribution of the excited states. We have noticed that the populations of the triple-well potentials are localized in the two outer wells.

These results have potential applications to explore more physical phenomena such as tunneling effect, and instantons dynamics.

[1]Department of Mathematics and Physics, California University, PA, USA
[2]Department of Applied Physics and Astronomy, University of Sharjah, Sharjah, UAE
[3]Institute of Quantum Science and Engineering, Texas A&M, TX, USA

## 1. Introduction

Multi-well is one of the most used model in molecular physics to describe equilibrium positions or molecular conformation [48] for multi-atomic molecules such as NH3 and ND3. In condensed matter physics, the quantum wells are modeled by a potential with double or multi-wells, in order to determine the transmission coefficients for the quantum transport [32], [54], and [23].

Quantum tunneling is known since the late 1920s, and investigating the phenomenon of tunneling dynamics in quantum models with triple-well potentials, is now at the heart of atomtronics devices such as tunneling diodes, transistors, scanning tunneling microscopes, and superconducting qubits for quantum computing, see [55], [27], and [50].

Tunneling is also crucial for several significant biological processes like photosynthesis, protein activation, self-repair of DNA, and enzymatic catalysis [31]. Another interesting area of research is tunneling effects in the brain, a review can be found in [1]. Triple-well is also used for optical waveguides design of light transfer [42], as well as to modelling quantum coherent transport of neutral atoms or electrons in multi-well traps [9].

For such types of studies exact solutions of wave functions and energies are needed. However, exact solutions of Schrödinger equations are limited to a small list of potentials and none of this list include double well or multi-well potentials. To the best of our knowledge, no exact whole spectrum of multi-well potentials are available. Recently, we have extended this list by two new potentials of finite bound states [8].

In this work, we present a new family of a triple-well potential where the eigenenergies are given explicitly and the wave functions are determined exactly. The potential is a sextic polynomial in its reduced form, but even though simple in its form, there is no known exact solution to this type of potentials.

Our study is based on the supersymmetric technique in quantum mechanics (SUSYQM), which is a well known method for generating exact energies and wavefunctions recursively.

More than three decades ago, it was conjectured in [4] and [15], that by applying SUSYQM method satisfying shape invariance property with linear parameter change function, the existing list of exactly solvable potentials is a complete list.

Here, we show that it is possible to generate exact solutions for potential partners with superpotentials composed of a linear combination of four independent elementary functions.

In section 2, we introduce supersymmetric approach in solving Schrödinger equation, followed by the shape invariance method. The new family of potentials and their graphs, their finite bound states eigenvalues, as well as their excited wavefunctions are introduced in section 3. We wrap up by a conclusion in section 4.

We would like to mention that, in the appendix, we summarize all known shape invariant exactly solvable potentials which are available in the literature, with their superpotentials, their partner potentials, and their bound state energies, in addition we complement them by the new solvable potentials proposed here and those proposed in our last paper [8].

## 2. Supersymmetry

We recall the definition of the supersymmetric potential partners $V_-(x)$, and $V_+(x)$, which are linked by the superpotential $W(x)$ by the following relations

$$V_\pm(x) = W^2(x) \pm W'(x) \tag{1}$$

Their associated Hamiltonian can be factorized as
$$H_- = A^\dagger A, \quad H_+ = AA^\dagger \tag{2}$$
where
$$A^\dagger = -\frac{d}{dx} + W(x), \quad A = \frac{d}{dx} + W(x) \tag{3}$$

It was shown that the eigenvalues and eigenfunctions of the two Hamiltonians $H_-$ and $H_+$ are related by (for $n = 0,1,2,...$)

$$E_0^{(-)} = 0, \quad E_n^{(+)} = E_{n+1}^{(-)}, \tag{4}$$

$$\Psi_n^{(+)} = \left(E_{n+1}^{(-)}\right)^{-1/2} A\Psi_{n+1}^{(-)} \tag{5}$$

$$\Psi_{n+1}^{(-)} = \left(E_n^{(+)}\right)^{-1/2} A^\dagger \Psi_n^{(+)}. \tag{6}$$

We focus in this paper on unbroken supersymmetry, the case where we assume the existence of a normalizable ground state $\Psi_0^{(-)}$ such that $A\Psi_0^{(-)} = 0$, so that $E_0^{(-)} = 0$.

The ground state wavefunction $\Psi_0^{(-)}$ can be expressed by

$$\Psi_0^{(-)} = N \; e^{-\int W(x)dx} \tag{7}$$

where $N$ is the normalized constant.

For a full discussion of unbroken and broken supersymmetries, please refer to [56], [57]. The above mentioned equations (energy and wave) allow us only to determine the spectrum of one potential from its partner and consequently, it permits to derive the whole spectrum of the supersymmetric partner potentials.

Two supersymmetric potential partners having shape invariance property if they satisfy the following condition:

$$V_-(x, a_1) + h(a_1) = V_+(x, a_0) + h(a_0) \tag{8}$$

where the parameter $a_1$ depends on $a_0$, i.e.

$$a_1 = f(a_0), \; a_2 = f^2(a_0)$$

and by recurrence

$$a_k = f^k(a_0), \text{where } a_0 \in \mathbb{R}^m$$

and where the function $f: \mathbb{R}^m \to \mathbb{R}^m$ is called a parameter change function, see [3], [11], [15], [17], [28].

By applying the shape invariance method, the parameters of the superpotentials are changed, but the partner potentials have similar shapes and only differ by a certain fixed constant.

It is still challenging to find solutions to the equation (8) of the shape invariant condition, except for the small list of the already known solvable potentials, such as, harmonic oscillator, Coulomb, Morse [46], Rosen-Morse [51], Eckart [24], Pöschl-Teller [49], Manning [45], Scarf [53], (see the whole list in the Appendix). We would like to mention that these Schrödinger equations can be transformed by some suitable change of variables to hypergeometric type differential equations, which can also be solved analytically or can be reduced to hypergeometric equation by using Nikiforv-Uvarov method. Recently, in [43], they gave the canonical transformations that connects one family to the other.

In [4],[5], and [15], it was suggested that there are no other shape invariant potentials beside the well known potentials when the parameters are related by a linear translation. It was noted in [11] that for the additive shape invariance potentials, they could generate the set of all known $\hbar$-independent shape invariant superpotentials, and then came to the conclusion that there are no others. However, these statements were limited to superpotentials written as linear combination of two functions.

In this paper, we have exhibited superpotentials as combination of four linearly independent elementary functions with a linear parameter change function.

In the appendix, a table that gives a summary of all known potentials with their superpotentials, partner potentials, and energies, followed by the new potentials. All the bound state energies, as well as, the expression of their wavefunctions are given explicitly.

From now on, we will consider the partner potential $V_-(x, a_0)$ as a shape-invariant potential. Therefore, the two potentials $V_-(x, a_1)$ and $V_+(x, a_0)$ have the same dependence on $x$, up to the change in their parameters, and their Hamiltonians $H_-(x, a_1)$ and $H_+(x, a_0)$ will only differ by a vertical shift given by $C(a_0) = h(a_1) - h(a_0)$,

$$V_+(x, a_0) = V_-(x, a_1) + C(a_0) \tag{9}$$

where the partner potentials are defined by

$$\begin{aligned} V_-(x, a_1) &= W^2(x, a_1) - W'(x, a_1), \\ V_+(x, a_0) &= W^2(x, a_0) + W'(x, a_0) \end{aligned} \tag{10}$$

Their common ground state wavefunction is given by

$$\Psi_0^{(+)}(x, a_0) = \Psi_0^{(-)}(x, a_1) \propto e^{-\int W(x, a_1)} \tag{11}$$

The first excited state $\Psi_1^{(-)}$ of $H_-(x, a_1)$ is given here, where we are omitting the normalization constant,

$$\Psi_1^{(-)}(x, a_0) = A^\dagger(x, a_0)\Psi_0^{(+)}(x, a_0) = A^\dagger(x, a_0)\Psi_0^{(-)}(x, a_1). \tag{12}$$

The eigenvalue associated to this Hamiltonian is

$$E_1^{(-)} = C(a_0) = h(a_1) - h(a_0) \tag{13}$$

The eigenvalues of the two Hamiltonians $H_+$ and $H_-$ have the same eigenvalues except for additional zero energy eigenvalue of the lower ladder Hamiltonian $H_-$. They are related by
$$\begin{aligned} E_0^{(-)} &= 0, \quad E_{n+1}^{(-)} = E_n^{(+)}, \\ \Psi_n^{(+)} &\propto A\Psi_{n+1}^{(-)}, A^\dagger\Psi_n^{(+)} \propto \Psi_{n+1}^{(-)}, \quad n = 0,1,2,... \end{aligned} \tag{14}$$

where we have iterated this procedure to construct a hierarchy of Hamiltonians

$$H_\pm^{(n)} = -\frac{d^2}{dx^2} + V_\pm(x, a_n) + \sum_{k=0}^{n-1} C(a_k) \tag{15}$$

and then derive the $n^{th}$ excited eigenfunction and eigenvalues by

$$\Psi_n^{(-)}(x, a_0) \propto A^\dagger(x, a_0)A^\dagger(x, a_1)...A^\dagger(x, a_n)\Psi_0^{(-)}(x, a_n) \tag{16}$$

$$E_0^{(-)} = 0,$$
$$E_n^{(-)} = \sum_{k=0}^{n-1} C(a_k) = \sum_{k=0}^{n-1} h(a_{k+1}) - h(a_k) \qquad (17)$$
$$= h(a_n) - h(a_0), \text{ for } n \geq 1.$$

where $a_k = f(f(\ldots f(a_0))) = f^k(a_0), k = 0,1,2,\ldots,n-1$.

Therefore, knowing the superpotential not only we know the potential, but also its ground state and from the algorithm above, the whole spectrum of the Hamiltonian $H$ ($H_+$ as well) can be derived by supersymmetry quantum mechanics method.

## 3. Exactly Solvable Symmetric Sextic Potential

The triple-well potential is now at the heart of extensive research field of atomtronics such as atomic transistor [55]. It is also applied in instantonic approaches analyzing nonperturbative aspects of quantum mechanical systems with degenerate vacua as well as tunneling effects [2].

We will introduce the new family of potentials constructed using shape invariance method where the superpotential

$$W(x, A, B, D, G) = Ax^3 + Bx - \frac{Dx}{1+Gx^2} \qquad (18)$$

is a four parameters linear combination of a cubic polynomial and a rational term and the corresponding potential partners $V_-$ and $V_+$, derived from Eq. (10), are written as a sextic polynomial in the form

$$V_-(x, A, B, D, G) = -B + 2G + \frac{1}{2}(2B^2 - 14GB - 7G^2)x^2 +$$
$$BG(2B + G)x^4 + \frac{1}{4}G^2(2B + G)^2 x^6 \qquad (19)$$

$$V_+(x, A, B, D, G) = B + 2G + \frac{1}{2}(2B^2 + 14GB - 7G^2)x^2 +$$
$$BG(2B - G)x^4 + \frac{1}{4}G^2(2B - G)^2 x^6 \qquad (20)$$

by using the fact that the potential satisfies the shape invariant condition and using the energy formula for such potential Eq. (10), with the parameters obtained recursively from Eq. (17) as follows:

$$D_0 = -2G_0, \quad A_0 = \frac{1}{2}(2B_0 - G_0)G_0, \qquad (21)$$

and the change of parameter function $f: \mathbb{R}^4 \to \mathbb{R}^4$ is defined by the linear relation in matrix form

$$\boldsymbol{a}_k = (-1)^k \boldsymbol{M}_0 \, \boldsymbol{a}_0 + 2(-1)^k k B_0 \boldsymbol{b}_0, \text{ for all } k \geq 1 \qquad (22)$$

where

$$\boldsymbol{a_0} = \begin{bmatrix} A_0 \\ B_0 \\ D_0 \\ G_0 \end{bmatrix}, \boldsymbol{b_0} = \begin{bmatrix} 0 \\ 0 \\ 2 \\ -1 \end{bmatrix}$$

$$\text{and } \boldsymbol{M_0} = \begin{bmatrix} 1 & 0 & 0 & 0 \\ 0 & 1 & 0 & 0 \\ 0 & 0 & -1 & 0 \\ 0 & 0 & 0 & 1 \end{bmatrix}, \tag{23}$$

Hence, the general expression of the bound energies $\varepsilon_k$, can be deduced using (17) and (22)

$$\varepsilon_k = 4(-1)^k B_0 + \frac{4(2B_0 - G_0 + k B_0)G_0}{-G_0 + 2B_0 k} \tag{24}$$

with odd and even eigenstates as follows

$$\begin{aligned} \varepsilon_0 &= 0, \quad \text{and for } n = 1, 2, \dots \\ \varepsilon_{2n-1} &= -4B_0 + \frac{4(2B_0 - G_0 + k B_0)G_0}{-G_0 + 2B_0 k} \end{aligned} \tag{25}$$

$$\varepsilon_{2n} = 4B_0 + \frac{4(2B_0 - G_0 + 2n B_0)G_0}{-G_0 + 4B_0 n}.$$

The proposed family of potentials are more general and contains triple-well as well as double-well and single-well.

In this paper, we will focus on the symmetric triple-well in the symmetric natural domain $(-\infty, \infty)$, where the potential has at most five relative extrema.

### 3.1. Triple-Well Potential

Now, using Eq. (21) and Eq. (24), let's express the potential $V_-(x, A_1, B_1, D_1, G_1)$, in terms of the two free parameters $(B_0, G_0)$ and let's assume

$$G_0 > 2B_0 > 0, \tag{26}$$

in order to satisfy the following physical constraint and avoid energy crossing,

$$E_{n+1} > E_n > 0 > \min(V_-) = -\varepsilon, \tag{27}$$

where, in this case $\varepsilon > 0$, then the absolute minimum value $(V_-)_{\min}$ of this potential is attained at $x = \pm x_0$, where $x_0$ is the positive real root of $V'_- = 0$, that is, a solution of the fourth degree equation

$$2(2B_0^2 + 14B_0 G_0 - 7G_0^2) + 8B_0 G_0(2B_0 - G_0)x^2 + 3G_0^2(2B_0 - G_0)^2 x^4 = 0 \tag{28}$$

and thus, if we assume $0 < 2B_0 < G_0 < \frac{7+3\sqrt{7}}{7} B_0$,
then the root $x_0$ will satisfy

$$x_0^2 = \frac{4B_0 - \sqrt{2(2B_0^2 - 42B_0G_0 + 21G_0^2)}}{3G_0(G_0 - 2B_0)} \tag{29}$$

and by evaluating $V_-$ at $x_0$, we obtain

$$\varepsilon = \frac{31B_0}{9} - 2G_0 - \frac{4B_0^3}{9G_0(2B_0 - 9G_0)} - \frac{1}{9}(2B_0^2 - 42B_0G_0 + 21G_0^2)x_0^2 \tag{30}$$

Now, let's consider the nonnegative triple-well potential $V$, by shifting the potential partner $V_-$ by the quantity $(V_-)_{\min}$, that is,

$$V = V_- - (V_-)_{\min} \tag{31}$$

We obtain the sextic potential expressed in terms of the two free parameters $(B_0, G_0)$

$$V(x, B_0, G_0) = \frac{2B_0(2B_0^2 + 14B_0G_0 - 7G_0^2)}{9(2B_0 - G_0)G_0} + \frac{1}{9}(2B_0^2 - 42B_0G_0 + 21G_0^2)x_0^2 +$$

$$\frac{1}{2}(2B_0^2 + 14B_0G_0 - 7G_0^2)x^2 + B_0G_0(2B_0 - G_0)x^4 + \frac{1}{4}G_0^2(2B_0 - G_0)^2 x^6 \tag{32}$$

Since the potential $V$ has a double root at $\pm x_0$, it is therefore possible to conveniently express this potential in a compact form

$$V(x, B_0, G_0) = \frac{1}{4}G_0^2(2B_0 - G_0)^2(x^2 - x_0^2)^2\left(x^2 + 2x_0^2 + \frac{2B_0^2 + 14B_0G_0 - 7G_0^2}{G_0^2(2B_0 - G_0)^2}\right) \tag{33}$$

The general expression of its bound energies $E_n$ is written in the form:

$E_0 = \varepsilon$, and for $n \geq 1$

$$E_{2n-1} = 2\left(-4B_0 + G_0 + 2B_0 n + \frac{(2B_0 - G_0)G_0}{-G_0 + 2B_0(-1 + 2n)}\right) + \varepsilon,$$

$$E_{2n} = 2\left(2B_0 + G_0 + 2B_0 n + \frac{(2B_0 - G_0)G_0}{-G_0 + 4B_0 n}\right) + \varepsilon. \tag{34}$$

From figure 1 of the triple-well potential $V$ and its bound state energies $E_n$ with their corresponding wavefunction $\Psi_n$, we note that the two excited states are localized in the two outer wells.

The relative gap energy ratio

$$\rho = \frac{\varepsilon}{E_2 - E_1} \tag{35}$$

which is the ratio of the difference between the zero point energy and the ground state $\varepsilon$, and the difference between the two consecutive excited states $(E_2 - E_1)$, is given by

$$\rho = \frac{\frac{31B_0}{9} - 2G_0 - \frac{4B_0^3}{9G_0(2B_0 - 9G_0)} - \frac{1}{9}(2B_0^2 - 42B_0G_0 + 21G_0^2)x_0^2}{12B_0 - 2G_0 + 2G_0\frac{G_0 - 2B_0}{G_0 - 4B_0}}, \tag{36}$$

and can be computed in the region obtained by imposing the physical constraints on the energies while still preserving the triple-well shape of the potential. We can show, under the parametric conditions above that the relative gap energy Eq. (36) satisfies

$$26.0765 < \rho, \tag{37}$$

for this triple-well potential, where the minimum value was attained on the boundary of the parametric region when $(B_0, G_0) = \left(1, \frac{7+3\sqrt{7}}{7}\right)$. Thus, $\rho$ is much bigger that $\rho_H$ of the harmonic oscillator which is exactly $\rho_H = \frac{1}{2}$.

Using Eq. (11) we obtain our ground state eigenfunction as a function of $(B_0, G_0)$ in the form

$$\Psi_0(x) = \frac{1}{1+G_0 x^2} e^{\frac{1}{8}(-4B_0+G_0(-2B_0+G_0)x^2)x^2} \tag{38}$$

and then using Eq. (12), we obtain the first excited state

$$\Psi_1(x) = -\frac{4(B_0-G_0)x}{1+G_0 x^2} e^{\frac{1}{8}(4B_0-G_0(-2B_0+G_0)x^2)x^2} \tag{39}$$

and then the second excited state

$$\Psi_2(x) = e^{\frac{1}{8}(4B_0-G_0(-2B_0+G_0)x^2)x^2} * \tag{40}$$
$$\frac{10B_0-4G_0+(12B_0^2-10B_0 G_0+G_0^2)x^2+4G_0(-3B_0+G_0)(-2B_0+G_0)x^4+3G_0^2(-2B_0+G_0)^2 x^6}{1+G_0 x^2}$$

For the ground state, we can observe from Eq. (38) that the coefficient $G_0(-2B_0 + G_0)$ of the exponential factor leading term is positive, and therefore $\Psi_0$ is divergent at infinity, however, its complement function $\widetilde{\Psi}_0 = \Psi_0 \int \frac{1}{\Psi_0^2}$, is an exponentially converging function. In order to show this, under the same parameter condition (26), let's define the function

$$g(x) = \frac{(1+G_0 x^2)^2}{x(-2B_0+G_0(-2B_0+G_0)x^2)} \tag{41}$$

This function is positive monotone increasing on $(\alpha, \infty)$, where $\alpha = \frac{3G_0+\sqrt{16B_0^2-8B_0 G_0+9G_0^2}}{2(G_0-2B_0)}$. Hence, for all $x \geq \alpha$,

$$\left|\widetilde{\Psi}_0(x)\right| = \frac{e^{\frac{1}{8}(-4B_0+G_0(-2B_0+G_0)x^2)x^2}}{1+G_0 x^2} \left|\int_\alpha^x g(s) d\left(e^{-\frac{2}{8}(-4B_0+G_0(-2B_0+G_0)s^2)s^2}\right)\right|$$

$$\leq \frac{e^{\frac{1}{8}(-4B_0+G_0(-2B_0+G_0)x^2)x^2}}{1+G_0 x^2} \int_\alpha^x |g(s)| d\left(e^{-\frac{2}{8}(-4B_0+G_0(-2B_0+G_0)s^2)s^2}\right)$$

$$\leq \frac{|g(x)|}{1+G_0 x^2} e^{\frac{-1}{8}(-4B_0+G_0(-2B_0+G_0)x^2)x^2} \tag{42}$$

$$\leq \frac{(1+G_0 x^2)}{x(-2B_0+G_0(-2B_0+G_0)x^2)} e^{\frac{-1}{8}(-4B_0+G_0(-2B_0+G_0)x^2)x^2}$$

and so, it is square integrable on the real line, and thus the ground state has a normalizable wavefunction. But, though the eigenfunction $\widetilde{\Psi}_0(x)$ is well defined and square integrable, it is not an allowed physical state, since it is concentrated mainly around the central well where its energy is below the minimum of the central well.

In this case of triple-well potentials, this family only admits a maximum of two excited states. Their eigenfunctions given by Eq. (39) and Eq. (40) are convergent, since this time the coefficient $-G_0(-2B_0+G_0)$ of the exponent factor leading term is negative. Therefore, the eigenfunctions $\Psi_1$ for state $n = 1$, and $\Psi_2$ for state $n = 2$, are both vanishing exponentially at infinity and square integrable, therefore normalizable.

From the ground state wave function expression (38), we can see that around the origin $\Psi_0$ is asymptotically equivalent to $e^{-\frac{(B_0+2G_0)}{2}x^2}$, which is a Gaussian. Therefore, around the central well, the potential seems to behave like the harmonic oscillator for the ground state, and we might be inclined to approximate it as a harmonic oscillator, however, as we have mentioned above, since $\rho$ is much bigger than $\rho_H$, this approximation breaks down. In addition, the effect of the two outer wells on all states is remarkable. Hence, the approximation of the central well with the harmonic oscillator is not valid.

It should be emphasized that the probability density $|\Psi_n(x)|^2$ is symmetric for all existing bound states as this can be explained by the fact that the potential itself is symmetric.

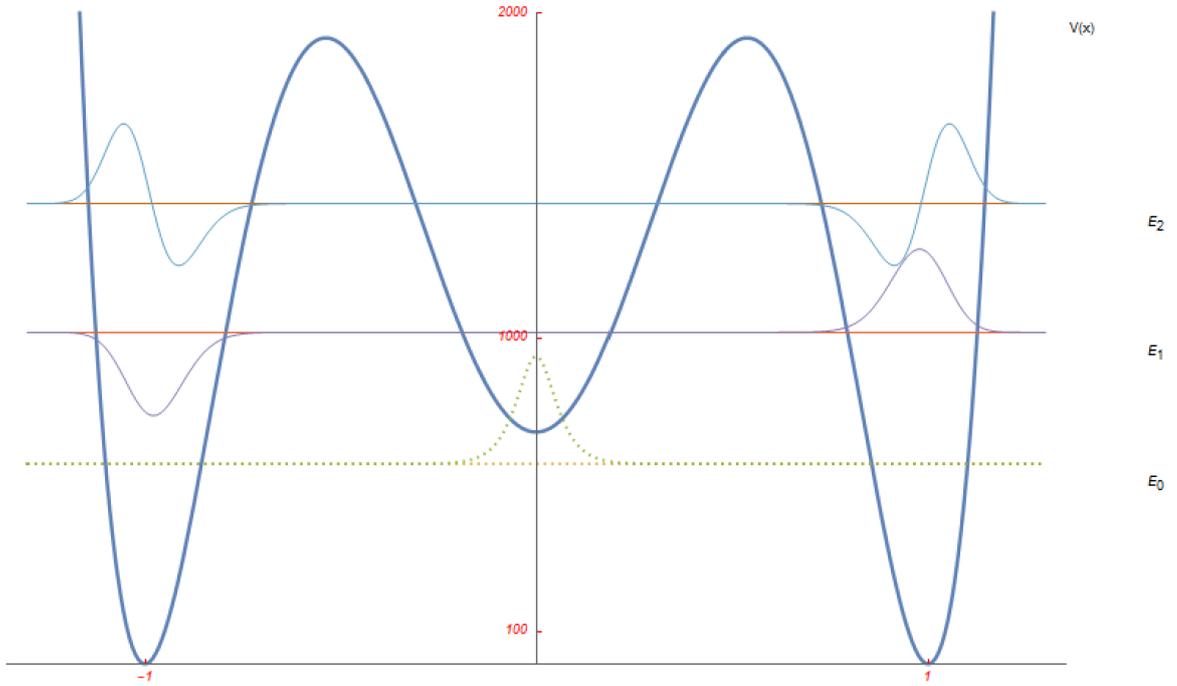

Figure 1: A symmetrict triple-well sextic potential V(x) given by Eq.(32) and $x_0^2$ by Eq.(29), here we have $x_0^2 = 1$, and parameters $B_0 = 100$, $G_0 = 100 + \sqrt{\frac{1}{3}\left(30407 + \sqrt{45649}\right)}$. The exact energy eigenvalues $E_n$, and their corresponding exact eigenfunctions $\psi_n(x)$ with the excited states localized in the outer wells. The ground state is seen as a fictitious state.

## 4. Conclusion

In this paper, we have discovered a new family of triple-well potentials, where we have expressed exactly their energies and eigenfunctions. As far as we know this is the first time that the whole spectrum of a triple-well potential is exactly obtained.

It was suggested since three decades ago that there are no superpotentials formed by more than two combination of functions verifying shape invariance, with a linear parameter change function $a_1 = a_0 +$ constant.

We have constructed a new family of superpotentials formed by a combination of three functions, two monomials and a rational function, satisfying the shape invariance property with linear parameter change function $a_1 = M_0 a_0 + b_0$, where $M_0$ is $4 \times 4$ matrix, $b_0$ is in $\mathbb{R}^4$. This new family of potentials will enrich the list of exactly solvable potentials. We have shown that the energy of the bound states are rational in the quantum number $n$. Furthermore, around the central well, approximating the central well by a harmonic oscillator as the usual procedure is not valid with this potential. The two outer wells affect noticeably the probability density distribution of the eigenstates.

From the plots of the triple-well potential and its bound state energies and the corresponding wave functions, we have noticed that the populations are localized in the two outer wells, which have been observed in the numerical plots in some recent works.

These results have potential applications for exploring more physical phenomena such as tunneling effect, and instantons dynamics.

**References**

[1] Adams, B., & Petruccione, F. Quantum effects in the brain: A review. AVS Quant. Sci., 2(2), 022901. (2020). doi:10.1116/1.5135170

[2] Alhendi, H.E., Lashin,E.I. Symmetric triple well with non-equivalent vacua: Instantonic approach. Mod. Phys. Lett. A Vol. 19, No. 28 (2004) 2103(2112)

[3] Bagchi, B.K. Supersymmetry in quantum and classical mechanics, Chapman&Hall, (2001).

[4] Barclay, D.T., Maxwell, C. J. Shape invariance and the SWKB series. Phys. Lett. A, Vol. 157, 6--7, (1991), pp. 357-360, doi.org/10.1016/0375-9601(91)90869-A.

[5] Barclay, D.T., Dutt, R, Gangopadhyaya, A., Khare, A., Pagnamenta, A. and Sukhatme, U. New Exactly Solvable Hamiltonians: Shape Invariance and Self-Similarity. Phys. Rev. A, 48, 4, (1993) pp. 2786-2797. doi:10.1103/PhysRevA.48.2786

[6] Balantekin, A.B, Aleixo, A.N. Algebraic Nature of shape-invariant and self-similar potentials. J. Phys. A: Math. Gen.(1999) 32 2785.

[7] Balantekin, A.B. Algebraic approach to shape invariance. Phys. Rev A., 57, 6 (1998) pp. 4188–4191. 10.1103/PhysRevA.57.4188.

[8] Benbourenane, J., Eleuch, H. Exactly solvable new classes of potentials with finite discrete energies. Results Phys., ISSN: 2211-3797, Vol: 17, (2020) Page: 103034

[9] Berezovoj, V. P., et al. Tunneling dynamics in exactly solvable models with triple-well potentials. J. Phys. A: Math. Theor.(2013) 46 065302

[10] Bougie, J., Gangopadhyaya, A., Mallow, J.V. Rasinariu, C. Generation of a novel exactly solvable potential. Phys. Lett. A, 379, (2015) pp. 2180-2183.

[11] Bougie, J., Gangopadhyaya, A., Mallow, J.V. Rasinariu, C. Supersymmetric quantum mechanics and solvable models. Symmetry, 4 (2012), pp 452-473.

[12] Bougie, J., Gangopadhyaya, A., Mallow, J.V. Generation of a Complete Set of Additive Shape-Invariant Potentials from an Euler Equation. Phys. Rev. Lett. 105, 210402 (2010)

[13] Chuan, C. Exactly solvable potentials and the concept of shape invariance. J. Phys.: Math. Gen. 24 (1991) L1165-L1174.

[14] Cooper, F., Khare, A., Sukhatme, U.P. Supersymmetry In Quantum Mechanics. World Sci. (2001)

[15] Cooper, F., Ginocchio, J. N., and Khare, A. Relationship between Supersymmetry and Solvable Potentials. Phys. Rev. D, 36, 8, (1987) pp. 2458-2473.

[16] Cooper, F., Khare, A. and Sukhatme, U. Supersymmetry and Quantum Mechanics. Phys. Rep. 251, 5-6, (1995) pp. 267-385. doi:10.1016/0370-1573(94)00080-M

[17] Dabrowska, J. W., Khare, A. and Sukhatme, U.P. Explicit Wavefunctions for Shape-Invariant Potentials by Operator Techniques. J. Phys. A: Math. Gen. 21, 4 (1988) pp. L195-L200. doi:10.1088/0305-4470/21/4/002

[18] Dong, S. H., Factorization Method in Quantum Mechanics, Springer (2007)

[19] Dunne, G. V., Sulejmanpasic, T., and Unsal, M. "Bions and Instantons in Triple-well and Multi-well Potentials." arXiv preprint arXiv:2001.10128 (2020).

[20] Dutt, R., Khare, A., Sukhatme, U.P. Supersymmetry, shape invariance, and exactly solvable potentials. Am. J. Phys., 56,(1988) pp. 163-168.

[21] Dutt, R., Khare, A., and Sukhatme, U.P. Exactness of Supersymmetric WKB Spectra for Shape-Invariant Potentials. Phys. Lett. B, 181, 3-4,(1986) pp. 295-298. doi:10.1016/0370-2693(86)90049-3

[22] Dutt, R., Gangopadhyaya, A., Khare, A., Pagnamenta, A., Sukhatme, U. Solvable quantum mechanical examples of broken supersymmetry. Phys. Lett. A, 174, 5-6,(1993) pp. 363-367. doi:10.1016/0375-9601(93)90191-2

[23] Dutta, S., et al. Management of the correlations of Ultracold Bosons in triple wells. New J. Phys. 21 053044 (2019)

[24] Eckart, C. The Penetration of a Potential Barrier by Electron. Phys. Rev. 35, 1303 (1930)

[25] C. D.J.F. Trends in Supersymmetric Quantum Mechanics. In: Kuru Ş, Negro J, Nieto L. (eds) Integrability, Supersymmetry and Coherent States. CRM Ser. Math. Phys. Springer, Cham. (2019) doi:10.1007/978-3-030-20087-9_2.

[26] Frank, W. M., Land, D. J., & Spector, R. M. Singular Potentials. Rev. Mod. Phy., 43(1), 36–98. . (1971) doi:10.1103/revmodphys.43.36

[27] Gajdacz, M., Opatrný, T., Das, K.K. An atomtronics transistor for quantum gates. Phys. Lett. A, 378, 28-29 (2014), pp. 1919-1924, doi:10.1016/j.physleta.(2014).04.043

[28] Gangopdhayaya, A., Mallow, J., Rasinariu, C. Sypersymmetry quantum: an introduction, Second edition., Singapore; Hackensack, NJ : World Sci. (2017)

[29] Gendenshtein, L.E. Derivation of exact spectra of the Schrödinger equation by means of supersymmetry, JETP Lett. 38 (1983) pp. 356–359

[30] Gendenshtein, L.E, and Krive, I.V. Supersymmetry in quantum mechanics. Sov. Phys. Usp. 28 (1985) pp. 645–666.

[31] Godbeer, A. D., Al-Khalili, J. S., and Stevenson, P. D. (2015). Modelling proton tunnelling in the adenine–thymine base pair. Phys. Chem. Chem. Phys. 17(19), 13034–13044. doi:10.1039/c5cp00472a

[32] Harrison, P., Valavanis, A. Quantum wells, wires and dots : theoretical and computational physics of semiconductor nanostructures, Wiley ( 2016) - 598 p.

[33] Infeld, L., Hull, T. E. The factorization method. Rev. Mod. Phys. 23,(1951) 21–68.

[34] Junker, G. Supersymmetry methods in quantum and statistical physics, IOP (2019).

[35] Kar, S., and P. K. Chattaraj. Tunneling and quantum localization in chaos-driven symmetric triple well potential: An approach using quantum theory of motion. Int. J. Quantum Chem. 118, no. 10 (2018): e25531

[36] Keung, W., Kovacs, E., Sukhatme, U.P. Supersymmetry and double-well potentials. Phys. Rev. Lett. 60, 1, (1988).

[37] Khare, A., Sukhatme, U. Phase equivalent potentials obtained from supersymmetry. J. Phys. A.; Math. Gen. 22, 2847, (1989).

[38] Khare, A., and Sukhatme, U.P. Scattering Amplitudes for Supersymmetric Shape-Invariant Potentials by Operator Methods. J. Phys. A: Math. Gen., 21, 9, (1988), pp. L501. doi:10.1088/0305-4470/21/9/005

[39] Khare, A., and Sukhatme, U.P. New Shape-Invariant Potentials in Supersymmetric Quantum Mechanics. J. Phys. A: Math. Gen., 26 (1993) pp. L901-L904. doi:10.1088/0305-4470/26/18/003

[40] Levai, G. A search for shape-invariant solvable potentials. J. Phys. A: Math. Gen, 22 (1989) pp. 689-702.

[41] Levai, G. Solvable potentials derived from supersymmetric quantum mechanics, pp. 107-126, in: Von Geramb H.V. (eds) Quantum Inversion theory applications. Lect. Notes Phys., 427, Springer (1994).

[42] Longhi, S., Della Valle, G., Ornigotti, G., M., and Laporta, P. Coherent tunneling by adiabatic passage in an optical waveguide system. Phys. Rev. B 76, 201101(R) (2007).

[43] Mallow, J. V., Gangopadhyaya, A., Bougie, J., & Rasinariu, C. Inter-relations between additive shape invariant superpotentials. Phys. Lett. A, 126129. (2019).

oi:10.1016/j.physleta.2019.126129

[44] Manning, M. F., Rosen, N.: Potential function of diatomic molecules. Phys. Rev. 44, 953 (1933)

[45] Manning, M. F. Exact Solutions of the Schrödinger Equation. Phys. Rev, 48(2), 161–164. (1935). doi:10.1103/physrev.48.161

[46] Morse, P. M. Diatomic molecules according to the wave mechanics II. Vibrational levels. Phys. Rev, 34(1), (1929) 57-64. doi:10.1103/physrev.34.57.

[47] Nicolai, H. Supersymmetry and system spin, J. Phys. A: Math. Gen. 9 1497 (1976) Doi:10.1088/0305-4470/9/9/010.

[48] Peacock-Lopez, E. Exact Solutions of the Quantum Double-Square-Well Potential. Chem. Educator (2006), 11, 383–393

[49] Pöschl, G., Teller, E. Bemerkungen zur quantenmechanik des anharmonischen oszillators. Z. Physik 83, 143-151(1933) doi:10.1007/BF01331132

[50] Ramos, R., Spierings, D., Racicot, I. et al. Measurement of the time spent by a tunnelling atom within the barrier region. Nature 583, 529–532 (2020). doi: 10.1038/s41586-020-2490-7

[51] Rosen, N., Morse, P. M. On the Vibrations of Polyatomic Molecules. Phys. Rev., 42(2), 210–217 (1932). doi:10.1103/physrev.42.210

[52] Roy, B., Roy, P., & Roychoudhury, R. On Solutions of Quantum Eigenvalue Problems: A Supersymmetric Approach. Fortschr. Phys./Progr. Phys., 39(3), 211--258 (1991). doi:10.1002/prop.2190390304

[53] Scarf, F. L. New Soluble Energy Band Problem. Phys. Rev., 112(4), 1137–1140 (1958). doi:10.1103/physrev.112.1137

[54] Schlagheck, P., et al, Transport and interaction blockade of cold bosonic atoms in a triple-well potential. New J. Phys. 12 065020 (2010)

[55] Stickney, J. A., Anderson, D.Z. and Zozulya, A.A.Transistorlike behavior of a Bose-Einstein condensate in a triple-well potential. Phys. Rev. A 75, 013608 ( 2007)

[56] Sukumar, C. V. Supersymmetric quantum mechanics in one-dimensional systems. J. Phys. A: Math. Gen. 18, 2917 (1985).

[57] Sukumar, C. V. Supersymmetry, factorization of the Schrödinger equation and a Hamiltonian hierarchy, J. Phys. A: Math. Gen. 18 L57 (1985).

[58] Witten, E. Dynamical breaking of supersymmetry. Nucl. Phys. B. 188, (1981) pp 513–54. Doi: 10.1016/0550-3213(81)90006-7 .

**Appendix**

We list here the all known solvable potentials and the newly discovered ones.

| Name | $a_0$ | $a_1 = f(a_0)$ | Superpotential $W$ | Partner Potential $V_-(x, a_1)$ | Energy $E_n^{(-)}$ |
|---|---|---|---|---|---|
| Harmonic | $(A,B)$ | $(A,B)$ | $Ax - B$ | $(Ax - B)^2 - A$ | $2An$ |
| Coulomb | $(A,B)$ | $\left(\frac{AB}{1+B}, 1+B\right)$ | $A - \frac{B}{r}$ | $A^2 + \frac{B(B-1)}{r^2} - \frac{2AB}{r}$ | $A^2 - \left(\frac{AB}{B+n}\right)^2$ |
| 3D-Oscillator | $(A,B)$ | $(A, 1+B)$ | $Ar - \frac{B}{r}$ | $A^2 r^2 + \frac{B(B-1)}{r^2} - A(2B+1)$ | $4An$ |
| Morse | $(A,B)$ | $(A-p, B)$ | $A - Be^{-px}$ | $A^2 + B^2 e^{-2px} - 2B(A + \frac{p}{2})e^{-px}$ | $A^2 - (A - pn)^2$ |
| Rosen-Morse I | $(A,B)$ | $(A-p, \frac{AB}{A-p})$ | $A\cot px + B$ | $-A^2 + B^2 + A(A+p)\csc^2 px + 2B\cot px$ | $B^2 - A^2 - \left[\frac{A^2 B^2}{(A-pn)^2} - (A - pn)^2\right]$ |
| Rosen-Morse II | $(A,B)$ | $(A-p, \frac{AB}{A-p})$ | $A\tanh px + B$ | $A^2 + B^2 - A(A+p)\operatorname{sech}^2 px + 2AB\tanh px$ | $A^2 + B^2 - \left[(A+pn)^2 + \frac{A^2 B^2}{(A+pn)^2}\right]$ |
| Eckart | $(A,B)$ | $(A+p, \frac{AB}{A-p})$ | $-A\coth px + B$ | $A^2 + B^2 + A(A-p)\operatorname{csch}^2 px - 2AB\coth p$ | $A^2 + B^2 - \left[(A+pn)^2 + \frac{A^2 B^2}{(A+pn)^2}\right]$ |
| Scarf I | $(A,B)$ | $(A+p, B)$ | $A\tan px - B\sec px$ | $-A^2 + (A^2 + B^2 - pA)\sec^2 px - B(2A-p)\sec px \tan px$ | $-A^2 + (A+pn)^2$ |
| Scarf II | $(A,B)$ | $(A-p, B)$ | $A\tanh px + B\operatorname{sech} px$ | $A^2 + (-A^2 + B^2 - pA)\operatorname{sech}^2 px - B(2A+p)\operatorname{sech} px \tanh px$ | $A^2 - (A - pn)^2$ |
| Poschl-Teller I | $(A,B)$ | $(A-p, B)$ | $A\coth px - B\operatorname{csch} px$ | $A^2 + (A^2 + B^2 + pA)\operatorname{csch}^2 px - B(2A+p)\coth px \operatorname{csch} p$ | $A^2 - (A - pn)^2$ |
| Poschl-Teller II | $(A,B)$ | $(A-p, B+p)$ | $A\tanh px - B\coth px$ | $(A-B)^2 + B(B-p)\operatorname{csch}^2 px - A(A+p)\operatorname{sech}^2 px$ | $(A-B)^2 - (A - B - 2pn)^2$ |
| New potentials | | | | | |
| Double angle | $(A,B)$ | $(A-p, B-2p)$ | $A\tanh px + B\tanh 2px$ | $-3p\left(A + B - \frac{7p}{2}\right) + \frac{p}{2}(A + B - 2p)(\tanh px - 2\tanh 2px)^2$ | $(A+B)^2 - (A - B - 3np)^2$ |
| Quadruple angle | $(A,B)$ | $(A-p, B-4p)$ | $A\tanh px + B\tanh 4px$ | $\frac{-49p}{12}\left(A + B - \frac{241p}{49}\right) + \frac{p}{12}(A + B - 4p)(\tanh px - 4\tanh 4px)^2$ | $(A+B)^2 - (A - B - (3n+2)p)^2$ |
| Sextic | $(A, B, D, G)$ | $(A, -B, 4B - D, \frac{1}{2}G(2B - G))$ | $Ax^3 + Bx - \frac{Dx}{1+Gx^2}$ | $-B + 2G + \frac{1}{2}(2B^2 - 14BG - 7G^2)x^2 - BG(2B+G)x^4 + \frac{1}{4}G^2(2B+G)^2 x^6$ | $4(-1)^n B + \frac{4(B - G + E\,n)G}{-G + 2E\,n}$ |

Table: Well known solvable potentials and the newly added potentials